\documentclass[aps,prb,preprint,showpacs,byrevtex]{revtex4}

\usepackage{times,mathptm}
\usepackage[dvips]{graphicx,color}

\begin{document}

\title{Physisorption of DNA nucleobases on h-BN and graphene: vdW-corrected DFT calculations}

\author{Jun-Ho Lee,$^{1}$ Yun-Ki Choi,$^{1}$ Hyun-Jung Kim,$^{1}$ Ralph H. Scheicher,$^{2}$ and Jun-Hyung Cho$^{1}$$^{*}$}
\affiliation{$^{1}$Department of Physics and Research Institute for Natural Sciences, Hanyang University, 17 Haengdang-Dong, Seongdong-Ku, Seoul 133-791, Korea\\
$^{2}$Division of Materials Theory, Department of Physics and Astronomy, {\AA}ngstr{\"o}m Laboratory, Uppsala University, Box 516, SE-751 20, Uppsala, Sweden}

\date{\today}

\begin{abstract}
We present a comparative study of DNA nucleobases [guanine (G), adenine (A), thymine (T), and cytosine (C)] adsorbed on hexagonal boron nitride (\textit{h}-BN) sheet and graphene, using local, semilocal, and van der Waals (vdW) energy-corrected density-functional theory (DFT) calculations. Intriguingly, despite the very different electronic properties of BN sheet and graphene, we find rather similar binding energies for the various nucleobase molecules when adsorbed on the two types of sheets. The calculated binding energies of the four nucleobases using the local, semilocal, and DFT+vdW schemes are in the range of 0.54 ${\sim}$ 0.75 eV, 0.06 ${\sim}$ 0.15 eV, and 0.93 ${\sim}$ 1.18 eV, respectively. In particular, the DFT+vdW scheme predicts not only a binding energy predominantly determined by vdW interactions between the base molecules and their substrates decreasing in the order of G$>$A$>$T$>$C, but also a very weak hybridization between the molecular levels of the nucleobases and the ${\pi}$-states of the BN sheet or graphene. This physisorption of G, A, T, and C on the BN sheet (graphene) induces a small interfacial dipole, giving rise to an energy shift in the work function by 0.11 (0.22), 0.09 (0.15), $-$0.05 (0.01), and 0.06 (0.13) eV, respectively.

\end{abstract}

\maketitle

\section{INTRODUCTION}

The interaction of DNA nucleobases with inert surfaces has attracted much attention because of its importance for molecular recognition and self-organization processes.~\cite{siw,liu,pos,pra,pra1,pra2,he,cho,min} Indeed, a great number of theoretical studies for the adsorption of the four DNA nucleobases [guanine (G), adenine (A), thymine (T), and cytosine (C)] on graphene have been performed to explore the binding mechanism and the relative binding strength of G, A, T, and C.~\cite{gow,ant,le,son,ort,ber} These essentially planar base-graphene model systems can simplify the structural complexity of a full three-dimensional DNA double-stranded or single-stranded polymer adsorbed on graphene, to provide a possibility for direct experimental characterization of the molecular interactions with graphene. According to previous density-functional theory (DFT) calculations carried out within the local density approximation (LDA),~\cite{gow} the binding energy of the nucleobases on graphene varies in the following hierarchy: G$>$A${\approx}$T${\approx}$C. Calculations using the more accurate second-order M{\o}ller-Plesset perturbation theory (MP2),~\cite{gow} as well as other calculations utilizing DFT methods including van der Waals (vdW) interactions,~\cite{ant,le} obtained binding energy strengths of nucleobases with graphene in the ordering of G$>$A$>$T$>$C, consistent with the single solute adsorption isotherm study at the graphite-water interface.~\cite{sow} The same trend was also confirmed using isothermal titration calorimetry.~\cite{var} In contrast to the relatively large number of studies involving graphene, there have been relatively few studies concentrating on the interaction of DNA with the heterogeneous boron nitride (BN) sheet, which has been successfully fabricated through the micromechanical cleavage method~\cite{pac} and the chemical-solution-derived method.~\cite{han}

The hexagonal BN sheet exhibits the same honeycomb lattice structure as graphene. However, the electronic properties of the two sheets are drastically different from each other: graphene is a gapless semimetal with the nonpolar nature of the homonuclear C$-$C bond, while the BN sheet is an insulator with the polar nature underlying the charge transfer between its constituent B and N atoms.~\cite{wata,hod} It is thus interesting to investigate and compare how such different electronic properties of the BN sheet and graphene affect the binding mechanism and the relative binding strength of the four nucleobases. Using LDA, Lin \textit{et al}. showed that the nucleobase molecules are bound to the BN sheet via a polar electrostatic interaction, indicating that the interaction in the base-BN systems is somewhat different from the ${\pi}$-${\pi}$ interaction between the nucleobases and the nonpolar graphene sheet.~\cite{lin} However, calculations based on LDA may not be reliable for this base-BN systems, since LDA cannot correctly describe the long-ranged vdW interactions.

In the present work, we investigate the adsorption of the four DNA nucleobases on the BN sheet and graphene, using vdW energy-corrected DFT (DFT+vdW scheme~\cite{tka}) calculations and, for comparison, LDA as well as generalized gradient approximation (GGA) calculations. Somewhat surprisingly, we find that the binding energy for a given base molecule is very similar on both the BN sheet and graphene. According to our analysis, it is revealed that despite the large differences in the individual atomic polarizabilities between the BN sheet and graphene, the vdW energy between adsorbed molecule and the BN sheet is nearly equal to the corresponding one in the base-graphene system, resulting in very similar contributions of vdW interactions to the binding energy. We find that the vdW interactions between base molecule and its substrate determine the sequence of the binding energy as G$>$A$>$T$>$C. Here, the contribution of vdW interactions to the binding energy amounts to ${\sim}$1 eV, indicating a strong physisorption. Our electronic-structure analysis shows that this physisorption induces an interfacial dipole between the base molecule and substrate due to charge rearrangement, thereby causing a work-function shift relative to isolated BN sheet or graphene.

\section{COMPUTATIONAL METHODS}

The DFT+vdW,~\cite{tka} GGA, and LDA calculations were performed using the FHI-aims~\cite{blu} code for an accurate, all-electron description based on numeric atom-centered orbitals, with ``tight" computational settings and accurate tier 2 basis sets. Calculations with larger tier 3 basis sets show that our binding energies are converged to within 0.01 eV. For the exchange-correlation energy, we employed the LDA functional of Ceperley-Alder~\cite{cep} and the GGA functional of Perdew-Burke-Ernzerhof (PBE).~\cite{per3} The DFT+vdW scheme was combined with the PBE functional, since PBE tends to underestimate the binding energy at organic/graphene interfaces.~\cite{cha} In this PBE+vdW scheme, the total energy is composed of the PBE energy ($E_{\rm PBE}$) and the vdW energy ($E_{\rm vdW}$) which is given by a sum of pairwise interatomic C$_6$$R^{-6}$ terms:
   \begin{equation}
     E_{\rm vdW}=-\frac{1}{2} \sum_{A,B} f_{\rm damp}(R_{AB},R_{A}^0,R_{B}^0)C_{6,AB}R_{AB}^{-6},
   \end{equation}
where $R_{AB}$ is the distance between atoms $A$ and $B$, $C_{6,AB}$ is the corresponding $C_6$ coefficient, $R_{A}^0$ and $R_{B}^0$ are the vdW radii, and $f_{\rm damp}(R_{AB},R_{A}^0,R_{B}^0)$ is a damping function eliminating the $R_{AB}^{-6}$ singularity at small distances. Here, the $C_6$ coefficients were computed in a first principles way from the PBE ground-state electron density using the recently developed DFT+vdW scheme.~\cite{tka}

The present base-substrate systems were modeled in a periodic slab geometry in which base molecules adsorb on one side of the BN sheet or graphene. In order to make the overlap with electronic states from neighboring base molecules negligibly small, we employed a 5${\times}$5 unit cell of the BN sheet (graphene) with a vacuum spacing of 60 ${\rm \AA}$ between adjacent BN (graphene) sheets. The base molecules were terminated at the cut bond to the sugar ring with a methyl group to generate an electronic environment in the nucleobase more closely resembling the environment in an extended DNA chain rather than that of just individual isolated bases by themselves and also to introduce a certain degree of steric hindrance when interacting with the substrate. The ${\bf k}$-space integration was done with a single ${\Gamma}$ point in the Brillouin zone of the 5${\times}$5 unit cell. All atoms were allowed to relax along the calculated forces until every residual force component amounted to less than 0.02 eV/{\AA}.

\section{RESULTS AND DISCUSSION}

We begin by describing the equilibrium structures of adsorbed G, A, T, and C on graphene obtained using the LDA, PBE, and PBE+vdW schemes. On the basis of previous theoretical calculations,~\cite{ort,gow,lin} we optimize the AB-stacking-like arrangements of all four nucleobases on graphene. Here, we use the optimized lattice constant of graphene as 2.445, 2.467, and 2.465 {\AA} (see Table I) for LDA, PBE, and PBE+vdW, respectively. The optimized AB-stacking-like structures for adsorbed G, A, T, and C on graphene are very similar to the corresponding ones on the BN sheet (see Fig. 1). Table II lists the vertical distance between the base molecule and graphene sheet, obtained using LDA, PBE, and PBE+vdW. We find that LDA, PBE, and PBE+vdW give the values of the vertical distance ranging 3.08${\sim}$3.17 {\AA}, 3.95${\sim}$4.02 {\AA}, and 3.26${\sim}$3.29 {\AA}, respectively. Thus, the PBE+vdW values of the vertical distance are seen to fall between the LDA and PBE ones.

\begin{figure}[ht]
\centering{ \includegraphics[width=7.0cm]{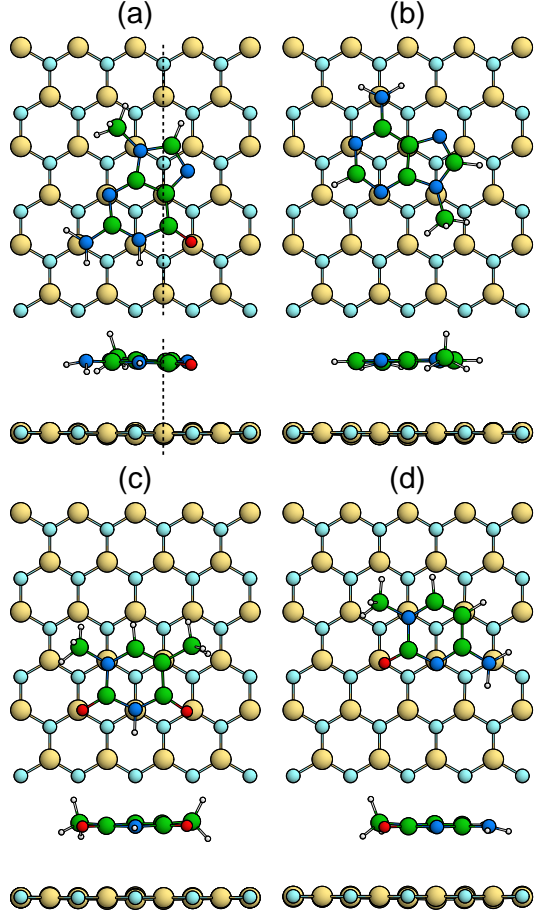} }
\caption{(Color online) Top and side views of the optimized structures of adsorbed (a) G, (b) A, (c) T, and (d) C on the BN sheet, respectively. The circles represent B, C, N, O, and H atoms with decreasing size. For distinction, N atoms on the BN sheet (nucleobases) are drawn in light (dark) color. The vertical plane along the dashed line is taken for the contour plot in Fig. 4.}
\end{figure}

\begin{table}[ht]
 \caption{
  Optimized lattice constant (in {\AA}) of the BN and graphene sheets obtained using LDA, PBE, and PBE+vdW, in comparison with previous theoretical results.
 }
 \begin{ruledtabular}
 \begin{tabular}{llccc}
          &         &  LDA   &  PBE    &  PBE+vdW \\  \hline
  BN       & This    & 2.489  & 2.513   &  2.510   \\
          & Ref. 32 & 2.49   &         &          \\
          & Ref. 33 &        & 2.511   &          \\
 graphene & This    & 2.445  & 2.467   &  2.465   \\
          & Ref. 30 & 2.445  &         &          \\
          & Ref. 31 &        & 2.47    &          \\
 \end{tabular}
 \end{ruledtabular}
 \end{table}

\begin{table}[ht]
 \caption{
 Calculated vertical distance (in {\AA}) between the base molecule and BN or graphene sheet, obtained using LDA, PBE, and PBE+vdW.
 }
 \begin{ruledtabular}
 \begin{tabular}{llcccc}
          &            &  G  &  A  &  T  &  C   \\  \hline
BN        &  LDA       & 3.03 & 3.04 & 3.08 & 3.12       \\
          &  PBE       & 3.80 & 3.94 & 3.96 & 4.04       \\
          &  PBE+vdW   & 3.21 & 3.25 & 3.27 & 3.24       \\
graphene  &  LDA       & 3.08 & 3.17 & 3.10 & 3.12       \\
          &  PBE       & 3.95 & 4.00 & 4.02 & 3.97       \\
          &  PBE+vdW   & 3.26 & 3.29 & 3.29 & 3.27       \\
 \end{tabular}
 \end{ruledtabular}
 \end{table}

The calculated binding energies $E_{\rm b}$ for the four base-graphene systems using LDA, PBE, and PBE+vdW are listed in Table III. We find that LDA (PBE; PBE+vdW) gives $E_{\rm b}$ = 0.72 (0.14; 1.18), 0.55 (0.06; 1.00), 0.54 (0.08; 0.95), and 0.56 (0.13; 0.93) eV for adsorbed G, A, T, and C on graphene, respectively. Thus, the binding sequence of the four nucleobases is G$>$A${\approx}$T${\approx}$C for LDA and G$>$A$>$T$>$C for PBE+vdW. This trend of the binding sequence for the four base-graphene systems agree well with previous theoretical results (see Table III),~\cite{gow,ant,le,son,ort} though the magnitudes of $E_{\rm b}$ vary depending on the employed implementations of various computational methods.~\cite{ref1,ref2} It is notable that PBE gives lower binding energies of less than ${\sim}$0.2 eV (Table III) and relatively larger vertical distances around ${\sim}$4.0 {\AA} (Table II), indicating that it does not correctly describe the ${\pi}-{\pi}$ stacking interactions between the base molecule and graphene. We also note that LDA predicts relatively lower binding energies yet smaller vertical distances compared with those obtained using PBE+vdW.

To assess the contribution of vdW interactions to the binding energy, we decompose the PBE+vdW binding energy into two parts computed from $E_{\rm PBE}$ and $E_{\rm vdW}$:
   \begin{equation}
    E_{\rm b} = E_{\rm b,PBE} + E_{\rm b,vdW},
   \end{equation}
where $E_{\rm b,PBE}$=$-$[$E_{\rm PBE}(\rm base/sub)-E_{\rm PBE}(\rm base)-E_{\rm PBE}(\rm sub)$] and $E_{\rm b,vdW}$=$-$[$E_{\rm vdW}(\rm base/sub)-E_{\rm vdW}(\rm base)-E_{\rm vdW}(\rm sub)$], in which ``sub'' stands for ``substrate''. Figure 2 shows $E_{\rm b,vdW}$ for the four base-graphene systems. We find that adsorbed G, A, T, and C on graphene have $E_{\rm b,vdW}$ ($E_{\rm b,PBE}$) = 1.18, 1,10, 1,02, and 0.94 (0.00, $-$0.10, $-$0.07, and $-$0.01) eV, yielding $E_{\rm b}$ = 1.18, 1.00, 0.95, and 0.93 eV, respectively. Thus, for all four nucleobases, the contribution of vdW interactions to the binding energy amounts to ${\sim}$1 eV, indicating strong physisorption. We note that the PBE+vdW binding energies for adsorbed G, A, T, and C on graphene are larger than the PBE ones by 1.04, 0.94, 0.87, and 0.80 eV, respectively (see Table III). These values of the binding energy difference between the PBE+vdW and PBE calculations are somewhat different from the corresponding ones of $E_{\rm b,vdW}$ due to the use of two different adsorption structures in the PBE+vdW and PBE calculations (see Table II). It is likely that the relatively shorter vertical distance in PBE+vdW compared with that in PBE gives rise to the Pauli repulsion between ${\pi}$ electrons of the nucleobase molecule and graphene, yielding a negative value for $E_{\rm b,PBE}$.

\begin{figure}[ht]
\centering{ \includegraphics[width=7.0cm]{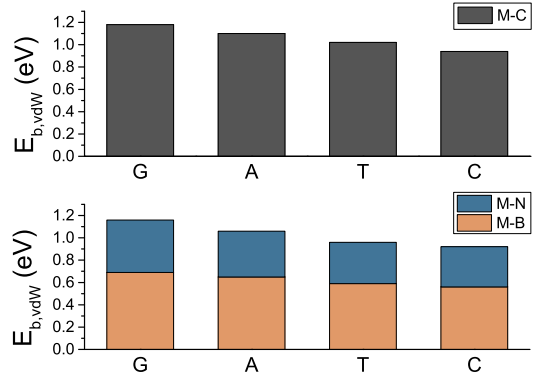} }
\caption{(Color online) Calculated $E_{\rm b,vdW}$ for adsorbed G, A, T, and C on the graphene (top panel) and BN sheets (bottom panel). M-C represents the value for the base-graphene systems. For the base-BN systems, the two components are decomposed: M-B (M-N) represents the component arising from the atomic pairs between the constituent atoms in the base molecule and the B (N) atom.}
\end{figure}

\begin{table}[ht]
 \caption{
 Calculated binding energies (in eV) of G, A, T, and C adsorbed on the BN and graphene sheets, in comparison with previous theoretical results.
 }
 \begin{ruledtabular}
 \begin{tabular}{lllcccc}
          &          &         &  G   &  A   &  T   &  C   \\  \hline
BN        &  LDA     & This    & 0.75 & 0.56 & 0.57 & 0.59       \\
          &          & Ref. 22 & 0.69 & 0.58 & 0.56 & 0.54    \\
          &  PBE     & This    & 0.15 & 0.07 & 0.08 & 0.13        \\
          &  PBE+vdW & This    & 1.18 & 1.01 & 0.94 & 0.93         \\
graphene  &  LDA     & This    & 0.72 & 0.55 & 0.54 & 0.56       \\
          &          & Ref. 10  & 0.61 & 0.49 & 0.49 & 0.49      \\
          &  PBE     & This    & 0.14 & 0.06 & 0.08 & 0.13         \\
          &  PBE+vdW & This    & 1.18 & 1.00 & 0.95 & 0.93         \\
          &          & Ref. 12 & 0.99 & 0.85 & 0.76 & 0.76         \\
          &  vdW-DF  & Ref. 12 & 0.74 & 0.63 & 0.60 & 0.58         \\
 \end{tabular}
 \end{ruledtabular}
 \end{table}

In Fig. 2, it is clearly seen that the binding energy sequence G$>$A$>$T$>$C is determined by vdW interactions between the base molecule and graphene. Since the vdW energy in the present PBE+vdW scheme is given by a sum of pairwise interatomic C$_6$$R^{-6}$ terms, we can say that the magnitude of the effective $C_{6,ij}$ coefficient between the base molecule and the C atom of graphene is in the same order of G$>$A$>$T$>$C as the binding energy, consistent with a previous theoretical study~\cite{gow} that the strength of the binding energy is governed by the polarizabilities of the base molecules which are in the order of G$>$A$>$T$>$C.

Next, we study the adsorption of four nucleobases on the BN sheet using the LDA, PBE, and PBE+vdW schemes. The optimized PBE+vdW structures for adsorbed G, A, T, and C on the BN sheet are respectively shown in the panels (a), (b), (c), and (d) of Fig. 1, showing the same AB-stacking-like arrangement as the case of the base-graphene systems. In Table II, we find that LDA, PBE, and PBE+vdW give the values of the vertical distance ranging 3.03${\sim}$3.12 {\AA}, 3.80${\sim}$4.04 {\AA}, and 3.21${\sim}$3.27 {\AA}, respectively. These values in the base-BN systems are slightly smaller than the corresponding ones in the base-graphene systems (see Table II). The calculated LDA (PBE; PBE+vdW) binding energies of G, A, T, and C on the BN sheet are 0.75 (0.15; 1.18), 0.56 (0.07; 1.01), 0.57 (0.08; 0.94), and 0.59 (0.13; 0.93) eV, respectively. It is seen that LDA predicts a slightly larger binding energy on the BN sheet, compared with that on graphene (see Table III). This may be attributed to a more electrostatic attraction between the nucleobase and BN sheet, due to the polar nature of the B$-$N bond. On the other hand, either PBE or PBE+vdW gives almost the same binding energy on the graphene and BN sheets, indicating that each base molecule binds to the two sheets with a nearly equal binding strength. Indeed, PBE+vdW predicts that adsorbed G, A, T, and C on the BN sheet have $E_{\rm b,vdW}$ = 1.16, 1.09, 0.99, and 0.93 eV, respectively. These values are very close to the corresponding ones (1.18, 1,10, 1,02, and 0.94 eV) on graphene. To understand this similarity of $E_{\rm b,vdW}$ on the two substrates, we decompose $E_{\rm b,vdW}$ obtained on the BN sheet into two components: i.e., one (the other) component arising from the atomic pairs between the constituent atoms in base molecule and the B (N) atom.~\cite{decomp} Such decompositions for the four base-BN systems are displayed in Fig. 2. We find that the ratio of magnitudes of the two components in all base-BN systems is about 8:5, indicating that the effective $C_{6,ij}$ coefficient between the base molecule (M) and the B atom is greater than that between M and the N atom by a factor of ${\sim}$1.6. Since the magnitudes of $E_{\rm b,vdW}$ in the base-graphene and base-BN systems are very close to each other, the value of ($C_{\rm 6,M-B}$ + $C_{\rm 6,M-N}$)/2 can be nearly equal to the value of $C_{\rm 6,M-C}$. The interesting conclusion from these results is that, despite the different bonding natures (i.e., non-polar and polar) of the graphene and BN sheets, the vdW interactions between each nucleobase molecule and the two substrates are close to each other, resulting in similar binding energies in the base-graphene and base-BN systems (see Table III). It is noteworthy that the in-plane polarizability of graphene and BN sheet would be very different, in the sense that the more localized electronic density of the polar network of B- and N-atom in the latter is more difficult to deform (from an energetic point of view) than in the more delocalized electronic density of the non-polar network of C-atoms in graphene. Despite this in-plane difference in polarizability, the main contribution to the vdW interaction with the ${\pi}$--${\pi}$ stacked nucleobases seems to arise however mainly from the eponymous ${\pi}$-orbitals belonging to the substrate (which are much closer in distance to the corresponding ${\pi}$-orbitals of the nucleobases). It is thus likely that the ${\pi}-{\pi}$ stacking interactions on the BN sheet will be very close to those on graphene, resulting in the overall quantitatively similar binding energy for a given nucleobase on both the BN sheet and graphene.

\begin{figure}[ht]
\centering{ \includegraphics[width=7.0cm]{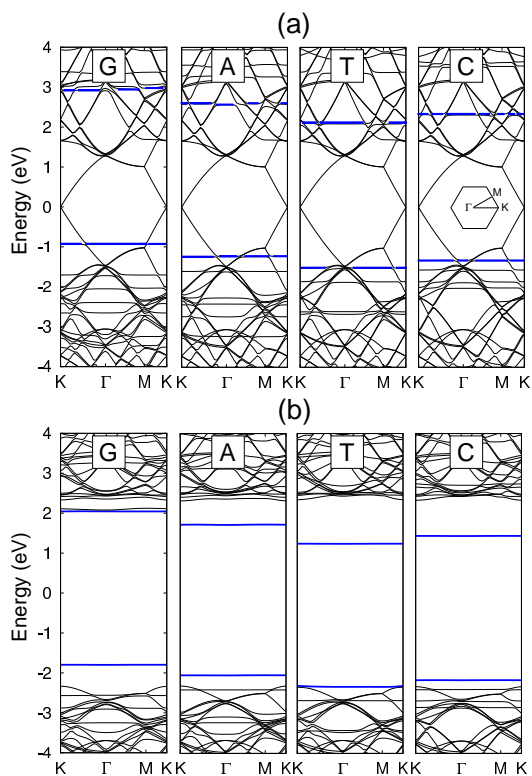} }
\caption{(Color online) Calculated band structures of adsorbed G, A, T, and C on the (a) graphene and (b) BN sheets. The band dispersions are plotted along the symmetry lines of the Brillouin zone of the 5${\times}$5 unit cell [see the inset in (a)]. The energy zero represents the Fermi level. For the base-BN systems, the Fermi level lies in the middle of the band gap of the BN sheet. For distinction, the highest occupied molecular orbital (HOMO) and lowest unoccupied state (LUMO) are drawn with thick lines.}
\end{figure}

Figure 3(a) and 3(b) show the calculated PBE+vdW band structures for adsorbed G, A, T, and C on the graphene and BN sheets, respectively. It is seen that the molecular orbitals of the four nucleobases hardly hybridize with the ${\pi}$ states of the graphene or BN sheet. We note that the HOMO of adsorbed G, A, T, and C on the graphene (BN) sheet locates at 0.93 (1.80), 1.23 (2.04), 1.53 (2.31), and 1.33 (2.17) eV below the Fermi level, respectively. Here, the HOMO positions follow the order of the ionization energies for the isolated base molecules G, A, T, and C which are calculated to be 5.39, 5.48. 5.79, and 5.60 eV, respectively. Our analysis of the Mulliken charges in the PBE+vdW calculations shows a very small charge transfer of less than 0.03 $e$ between any of the four nucleobases and the graphene or BN sheet. To see the rearrangement of charge at the base-substrate interface, we calculate the charge density difference defined as
   \begin{equation}
    {\Delta}{\rho} = {\rho}_{\rm M/sub} - ({\rho}_{\rm M} + {\rho}_{\rm sub}),
   \end{equation}
where ${\rho}_{\rm M/sub}$, ${\rho}_{\rm M}$, and ${\rho}_{\rm sub}$ denote the charge densities of the base-substrate system and its separated systems, i.e., isolated layer of base molecules and clean substrate, respectively. Figure 4(a) and 4(b) show ${\Delta}{\rho}$ for adsorbed G on the graphene and BN substrates, respectively. It is seen that the two base-substrate systems involve charge rearrangement at the interface. This charge rearrangement gives rise to an interfacial dipole, thereby causing the work-function shift.~\cite{wan,zie,hon} The calculated work-function shifts of all base-substrate systems are listed in Table IV. We find that, upon adsorption of G, A, T, and C on the graphene (BN) sheet, the work function increases relative to the value of 4.24 (3.48) eV at the isolated graphene (BN) sheet by 0.22 (0.11), 0.15 (0.09), 0.01 ($-$0.05), and 0.13 (0.06) eV per molecule within a 5${\times}$5 unit cell, respectively. We note that the work function of isolated graphene is calculated to be 4.24 eV, which is somewhat underestimated compared to the experimental~\cite{osh} value (${\sim}$4.6 eV) for graphite but consistent with other PBE calculations.~\cite{gio} The work-function shift ${\Delta}W$ can be correlated with the induced interfacial dipole ${\Delta}p$ by a simple electrostatics relation ${\Delta}W$ = $e$${\Delta}p$/(${\epsilon}_{0}A$),~\cite{leu} where $A$ is the area of 5${\times}$5 unit cell. Using this relation, we estimate ${\Delta}p$ as 0.065 (0.032), 0.045 (0.028), 0.002 ($-$0.015), and 0.040 (0.019) $e${\AA} for adsorbed G, A, T, and C on the graphene (BN) sheet, respectively. Note that ${\Delta}p$ corresponds to the normal component of induced dipole moment directed from base molecule to substrate. On the basis of our PBE+vdW results, we can say that, although the binding mechanism between the nucleobases and the graphene or BN sheet is driven by the vdW interactions, the interfacial dipole is induced upon adsorption to yield the work-function shift in the order of G$>$A$>$C$>$T. Here, we note that the values of ${\Delta}p$ for adsorbed T on the graphene and BN sheets are close to zero, possibly because of the cancelation of inhomogeneous interfacial dipole moments.

\begin{figure}[ht]
\centering{ \includegraphics[width=7.0cm]{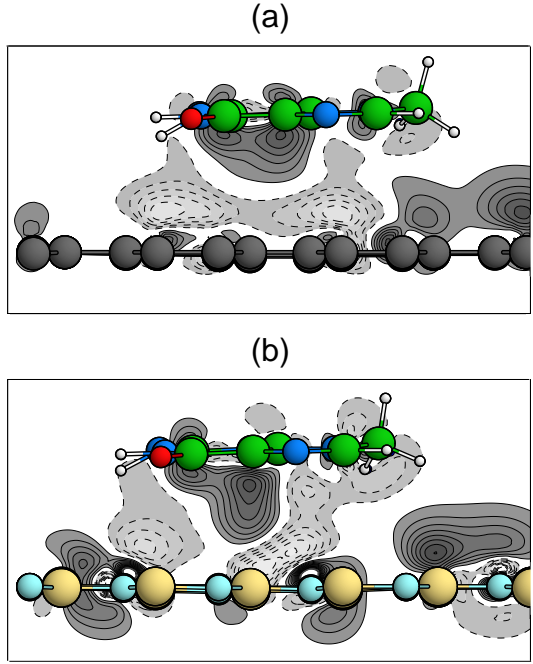} }
\caption{(Color online) Calculated charge density difference ${\Delta}{\rho}$ for adsorbed
G on the (a) graphene and (b) BN substrates. The contour plot in (a) is drawn in the vertical
plane along the dashed line in Fig. 1(a). The same vertical plane is also used in (b). The first solid (dashed) line is at
0.0004 ($-$0.0004) $e$/{\AA}$^3$ with
spacings of 0.0004 $e$/{\AA}$^3$.}
\end{figure}

\begin{table}[ht]
 \caption{
 Calculated work-function shift (in eV) upon physisorption of nucleobases G, A, T, and C on the BN and graphene sheets. The reference work functions of pristine BN and graphene sheets are given in the last column (in eV).
 }
 \begin{ruledtabular}
 \begin{tabular}{lccccc}
          &    G  &  A  &  T  &  C  & pristine  \\  \hline
BN        &   $+$0.11 &  $+$0.09 & $-$0.05 & $+$0.06 & 3.48   \\
graphene  &   $+$0.22 &  $+$0.15 & $+$0.01 & $+$0.13 & 4.24  \\
 \end{tabular}
 \end{ruledtabular}
 \end{table}

\section{SUMMARY}

We have investigated the adsorption of the four DNA nucleobases on the BN sheet and on graphene, using the LDA, PBE, and PBE+vdW schemes. The calculated binding energies of the four nucleobases on the two different substrates were predicted in the order of PBE+vdW $>$ LDA $>$ PBE. We found that the vdW interactions between each base molecule and the two sheets are very close to each other, giving rise to similar binding energies in the base-BN and base-graphene systems. Here, the magnitudes of the vdW interactions range from 0.9 to 1.2 eV, indicating a strong physisorption. We also found that the variation of vdW interactions depending on the base molecules determines the sequence of the binding energy as G$>$A$>$T$>$C, following the hierarchy of polarizabilities of the four DNA nucleobases. Our analysis shows that this physisorption induces an interfacial dipole between the base molecule and the substrate, leading to a small change in the work function relative to isolated graphene and BN sheets.

\noindent {\bf Acknowledgement.}
This work was supported by a National Research Foundation of Korea (NRF) grant funded by the Korean Government (MEST) (no. 2011-0031286), the Sweden-Korea Research Cooperation Programme of the Swedish Foundation for International Cooperation in Research and Higher Education (STINT) through Grant no. 2011/036, and KISTI supercomputing center through the strategic support program for the supercomputing application research (KSC-2012-C3-18). J.H.L. acknowledges support from the TJ Park Foundation.

\noindent $^{*}$ Corresponding author: chojh@hanyang.ac.kr


\begin{thebibliography}{99}

\bibitem{siw} Z. S. Siwy and M. Davenport, Nat. Nanotechnol. {\bf 5}, 697 (2010).
\bibitem{liu} M. Liu, H. Zhao, S. Chen, H. Yu, and X. Quan, Chem. Commun. {\bf 48}, 564 (2012).
\bibitem{pos} H. W. Ch. Postma, Nano Lett. {\bf 10}, 420 (2010).
\bibitem{he} F. Lu, F. Wang, L. Cao, C. Y. Kong, and X. Huang,
Nanosci. Nanotechnol. Lett. {\bf 4}, 949 (2012).
\bibitem{pra} R. H. Scheicher, A. Grigoriev, and R. Ahuja, J. Mater.
Sci. {\bf 47}, 7439 (2012).
\bibitem{pra1}  S. Mukhopadhyay, R. H. Scheicher, R. Pandey, and S. P. Karna,
J. Phys. Chem. Lett. {\bf 2}, 2442 (2011).
\bibitem{pra2} S. Mukhopadhyay, S. Gowtham, R. H. Scheicher, R. Pandey, and S. P. Karna, Nanotechnology {\bf 21}, 165703 (2010).
\bibitem{cho} T. Ahmed, S. Kilina, T. Das, J. T. Haraldsen, J. J. Rehr, and A. V. Balatsky,
Nano Lett. {\bf 12}, 927 (2012).
\bibitem{min} S. K. Min, W. Y. Kim, Y. Cho, and K. S. Kim, Nat. Nanotechnol. {\bf 6}, 162 (2011).

\bibitem{gow} S. Gowtham, R. H. Scheicher, R. Ahuja, R. Pandey, and S. P. Karna, Phys. Rev. B {\bf 76}, 033401 (2007).
\bibitem{ant} J. Antony and S. Grimme, Phys. Chem. Chem. Phys. {\bf 10}, 2722 (2008).
\bibitem{le} D. Le, A. Kara, E. Schr\"{o}der, P. Hyldgaard, and T. S. Rahman, J. Phys.: Condens. Matter {\bf 24}, 424210 (2012).

\bibitem{son} B. Song, G. Cuniberti, S. Sanvito, and H. Fang, Appl. Phys. Lett. {\bf 100}, 063101 (2012).
\bibitem{ort} F. Ortmann, W. G. Schmidt, and F. Bechstedt, Phys. Rev. Lett. {\bf 95}, 186101 (2005).
\bibitem{ber} K. Berland, S. D. Chakarova-K\"{a}ck, V. R. Cooper, D. C. Langreth, and E. Schr\"{o}der, J. Phys.: Condens. Matter {\bf 23}, 135001 (2011).

\bibitem{sow} S. J. Sowerby, C. A. Cohn, W. M. Heckl, and N. G. Holm, Proc. Natl. Acad. Sci. {\bf 98}, 820 (2001).
\bibitem{var} N. Varghese, U. Mogera, A. Govindaraj, A. Das, P. K. Maiti, A. K. Sood, and C. N. R. Rao, ChemPhysChem. {\bf 10}, 206 (2009).
\bibitem{pac} D. Pacil\'{e}, J. C. Meyer, C. \"{O}. Girit, and A. Zettl, Appl. Phys. Lett. {\bf 92}, 133107 (2008).
\bibitem{han} W.-Q. Han, L. Wu, Y. Zhu, K. Watanabe, and T. Taniguchi,
Appl. Phys. Lett. {\bf 93}, 223103 (2008).
\bibitem{wata} K. Watanabe, T. Taniguchi and H. Kanda, Nat. Mater. {\bf 3}, 404 (2004).
\bibitem{hod} O. Hod, J. Chem. Theory Comput. {\bf 8}, 1360 (2012).
\bibitem{lin} Q. Lin, X. Zou, G. Zhou, R. Liu, J. Wu, J. Li, and W. Duan, Phys. Chem. Chem. Phys. {\bf 13}, 12225 (2011).

\bibitem{tka} A. Tkatchenko and M. Scheffler, Phys. Rev. Lett. {\bf 102}, 073005 (2009).
\bibitem{blu} V. Blum, R. Gehrke, F. Hanke, P. Havu, V. Havu, X. Ren, K. Reuter, and M. Scheffler, Comput. Phys. Commun. {\bf 180}, 2175 (2009).
\bibitem{cep} D. M. Ceperley and B. J. Alder, Phys. Rev. Lett. {\bf 45}, 566 (1980).
\bibitem{per3} J. P. Perdew, K. Burke, and M. Ernzerhof, Phys. Rev. Lett. {\bf 77}, 3865 (1996); {\bf 78}, 1396(E) (1997).
\bibitem{cha} S. D. Chakarova-K\"{a}ck, E. Schr\"{o}der, B. I. Lundqvist, D. C. Langreth, Phys. Rev. Lett. {\bf 96}, 146107 (2006); D. C. Langreth, B. I. Lundqvist, S. D. Chakarova-K\"{a}ck, V. R. Cooper, M. Dion, P. Hyldgaard, A. Kelkkanen, J. Kleis, L. Kong, S. Li, P. G. Moses, E. Murray, A. Puzder, H. Rydberg, E. Schr\"{o}der, and T. Thonhauser, J. Phys.: Condens. Matter {\bf 21}, 084203 (2009).

\bibitem{ref1} Gowtham \textit{et al}. (Ref. 10) calculated the binding energy for adsorbed nucleobases on graphene using the LDA. The binding energy of the system consisting of the nucleobase and the graphene sheet is taken as the energy of the equilibrium configuration with reference to the asymptotic limit obtained by varying the distance between the base and the graphene sheet in the $z$ direction (Table II). For this equilibrium configuration, we obtained the binding energy close to the value obtained by Gowtham \textit{et al}.
\bibitem{ref2} Our PBE+vdW values of $E_{\rm b}$ for the four base-graphene systems are relatively larger compared with those of a previous~\cite{le} PBE+vdW calculation (see Table III). This may be due to the fact that the present calculation uses the base molecules with a methyl group attached, possibly giving rise to a larger binding energy because of an increase in the number of constituent atoms.

\bibitem{gio2} G. Giovannetti, P. A. Khomyakov, G. Brocks, P. J. Kelly, and J. van den Brink, Phys. Rev. B {\bf 76}, 073103 (2007). 
\bibitem{chan} K. T. Chan, J. B. Neaton, and M. L. Cohen, Phys. Rev. B {\bf 77}, 235430 (2008).  
\bibitem{sac} B. Sachs, T. O. Wehling, M. I. Katsnelson, and A. I. Lichtenstein, Phys. Rev. B {\bf 84}, 195414 (2011).  
\bibitem{pen} Q. Peng, W. Ji, and S. De, Comp. Mater. Sci. {\bf 56}, 11 (2012). 

\bibitem{decomp} This decomposition is allowed for the base-BN system because the sum of the vdW energies arising from the atomic pairs in base molecule and substrate is approximately equal to that arising from the corresponding ones in isolated base molecule and isolated substrate.

\bibitem{wan} B. Wang, S. G\"{u}nther, J. Wintterlin, and M.-L. Bocquet, New J. Phys. {\bf 12}, 043041 (2010).
\bibitem{zie} D. Ziegler, P. Gava, J. G\"{u}ttinger, F. Molitor, L. Wirtz, M. Lazzeri, A. M. Saitta, A. Stemmer, F. Mauri, and C. Stampfer, Phys. Rev. B {\bf 83}, 235434 (2011).
\bibitem{hon} S.-Y. Hong, P.-C. Yeh, J. I. Dadap, and R. M. Osgood, Jr., ACS Nano. {\bf 6}, 10622 (2012).
\bibitem{osh} C. Oshima and A. Nagashima, J. Phys.: Condens. Matter {\bf 9}, 1 (1997).
\bibitem{gio} G. Giovannetti, P. A. Khomyakov, G. Brocks, V. M. Karpan, J. van den Brink, and P. J. Kelly, Phys. Rev. Lett. {\bf 101}, 026803 (2008).
\bibitem{leu} T. C. Leung, C. L. Kao, W. S. Su, Y. J. Feng, and C. T. Chan, Phys. Rev. B {\bf 68}, 195408 (2003).

\end{thebibliography}
\end{document}